\documentclass[amssymb,aps,prd,reprint,nofootinbib,showpacs,superscriptaddress,groupedaddress]{revtex4-2}  
\usepackage{latexsym}
\usepackage{amsmath}
\usepackage[utf8x]{inputenc}
\usepackage[T1]{fontenc}
\usepackage{subdepth}
\usepackage{color}

\begin{document}
\def\calr{{\cal R}}
\def\GB{{\hat{\cal{G}}}}
\newcommand{\RR}{(*R*)}
\def\half{\textstyle{1\over2}}
\def\third{\textstyle{1\over3}}
\def\quarter{\textstyle{1\over4}}
\def\nn{\nonumber}
\def\Vg{V_{george}}
\def\Vr{V_{ringo}}
\def\Vj{V_{john}}
\def\Vp{V_{paul}}
\def\tap{{}^+\tau}
\def\tam{{}^-\tau}
\title{BTZ entropy from topological M-theory}
\author{Javier Chagoya$^1$}
\email{javier.chagoya@fisica.uaz.edu.mx}
\author{Graciela Reyes-Ahumada$^{2}$}
\email{grace@cimat.mx}
\author{M. Sabido$^{3}$}
\email{msabido@fisica.ugto.mx}
\affiliation{
$^{1}$Unidad Acad\'emica de F\'isica, Universidad Aut\'onoma de Zacatecas,
98060, M\'exico.\\
$^{2}$Conacyt - Unidad Académica de Matemáticas, Universidad Aut\'onoma de Zacatecas,
98060, M\'exico.\\
$^{3}$Departamento  de F\'{\i}sica de la Universidad de Guanajuato, A.P. E-143, C.P. 37150, Le\'on, Guanajuato, M\'exico.
 }%
\begin{abstract}
By determining the relation between topological M-theory and the Chern-Simons actions 
for a gauge field constructed from the Lie algebra of either $SL(2,\mathbb R)\times SL(2,\mathbb R)$
or $SL(2,\mathbb C)\times SL(2,\mathbb C)$, depending on the sign of the space-time curvature,
we show that the standard and exotic actions of 3-dimensional
gravity can be recovered from topological M-theory. With this result, we provide
a concrete realisation of a conjecture by Dijkgraaf et al.\!  stating that the partition function of topological M-theory is equivalent to the partition function of a black hole in 
a related theory. We do this for the standard and exotic BTZ black holes in 3-dimensional gravity. 

 \end{abstract}

\maketitle

 One of the most useful tools to understand the gravitational interaction, is three dimensional gravity. Not only has 2+1 gravity  been quantized, it has other remarkable features that are of great value as a guide to understand the foundations of gravity. Some of these features  can be easily derived from the fact that it can be written as a Chern-Simons action \cite{Witten:1988hc}. And although 2+1 gravity is topological and therefore might seem physically unrealistic (it lacks propagating degrees of freedom), there is a black hole solution, known as the BTZ black hole \cite{Banados1992}. The BTZ solution is asymptotically anti-de Sitter and has no singularity, but it has many of the features of the Kerr black hole, it has an event horizon and an inner horizon for the rotating case and thermodynamic properties analogous to 4 dimensional black holes. {Interestingly, the BTZ solution solves any 2+1 gravity model that admits anti-de Sitter   vacuum, and the mass and angular momentum  to some linear combination of the parameters}. When the role of mass and angular momentum is reversed, the resulting black hole is known as an exotic BTZ black hole. The  entropy of the BTZ black hole is in agreement with Hawking-Bekenstein entropy, but for the exotic case the entropy is related to the inner horizon. This appalling contradiction was resolved in \cite{Townsend2013} by considering that the BTZ is a solution to the standard action and the exotic BTZ is a solution to the exotic action, and therefore  the entropy must be given by, 
\begin{equation}
S=\frac{\pi}{2G}(\alpha r_+ + \gamma r _-).
\end{equation}
The standard and exotic actions, are the two independent actions that are derived in the Chern-Simons formulation of 2+1 gravity \cite{Witten:1988hc,PhysRevD.61.085022}.

The description of the gravitational field in terms of gauge fields or $p$-forms  has been 
continuously developed. 
In these theories the metric does not appear explicitly but it
is reconstructed from the dynamical fields under consideration. These descriptions 
are referred to as \emph{form theories of gravity}.
Some of these form theories, including Chern-Simons (CS) three dimensional (3d) gravity and the 
A and B models of topological strings can be unified in a seven dimensional space-time, $X$, through the 
topological M-theory (TMT) proposed by Dijkgraaf et al. \cite{Dijkgraaf:2004te}. Essential in
this theory is the \emph{volume form}, $V$, constructed from  an invariant $p$-form whose existence 
is characteristic of special holonomy manifolds. The study of manifolds
 admitting  stable non-degenerate forms is an interesting topic by itself, see for
 example \cite{2007HVLe} for a 
classification of all stable forms on $\mathbb R^n$. In particular, for seven dimensions, there are two 
non-trivial $p$-forms invariant under the holonomy group $G_2$, one of which is a 3-form and the
 other a 4-form. 
The same is true for the stable $p$-forms invariant under the dual group $\tilde G_2$. Using
the 3-form $\Phi$ with holonomy in $G_2$, Dijkgraaf et al. showed that the equations of motion for 2+1
gravity are recovered under a convenient partition of $X$. However, it is
 known that for non-vanishing cosmological constant, $\lambda$,  there are two classically equivalent 
actions to describe gravity in 2+1 dimensions,
 known as \emph{standard} and \emph{exotic} actions  \cite{Witten:1988hc}. 
In \cite{Chagoya_2018}, the authors obtain these actions from TMT. 
To write down the standard and exotic actions they first show how to obtain the CS actions for 2+1
gravity. This result opens up the possibility to apply the formalism and ideas of TMT to several models of 2+1 gravity that are built in terms of the CS actions.
Using this result we present a concrete realisation of a conjecture that states that the partition
function of a theory with an action 
defined by the Hitchin volume functional, is related to the partition function of a BPS black hole in
the gravitational theory allowed by the $p$-forms used to construct the volume functional   \cite{Dijkgraaf:2004te}.
Since we deal with 3d gravity, the relevant black hole solution will be the BTZ space-time \cite{Banados1992}. As a side result  we shed some light 
on the proposal in \cite{Townsend2013} for the entropy of BTZ black holes.

In~\cite{Ooguri:2004zv}, it is conjectured that the partition function
of a 4d BPS black hole is related to the topological string partition function by
\begin{equation}
    Z_{BH} = |Z_{top}|^2\,.\label{eq:ooguri}
\end{equation}
Furthermore, it is pointed out that the topological partition function can
be interpreted as a wave function, this interpretation comes from
\cite{witten1993quantum}. Thus, the conjecture above becomes
\begin{equation}
    Z_{BH} = |Z_{top}|^2 = |\psi|^2\,.
\end{equation}

A similar proposal exists in TMT \cite{Dijkgraaf:2004te}, where the partition function $Z_H$ of a 6d theory (contained within TMT and constructed from a volume form) is associated to a Wigner function arising from
the B model of topological strings. 

Here we present a realization of these ideas, but in the context
of 3d gravity. As shown in~\cite{Dijkgraaf:2004te} at the
level of the equations of motion and in \cite{Chagoya_2018} at the level
of the action, 3d gravity is contained in TMT as a particular splitting of the 7d manifold.
In order to give a concrete example of the relation between
$Z_H$ and the black hole entropy we consider an extremal BTZ
black hole, compute its volume form in terms of the
2+1 dimensional standard and exotic actions for gravity, then we obtain $Z_H$,
and finally we compare it to the norm of the wave function
for the same black hole \cite{Vaz2008}.


The organization of this work is as follows. First, we review and formalize the
derivation of the standard and exotic actions for 2+1 gravity from
TMT and construct the topological partition function. Then, we also review the BTZ black hole solutions and its partition function obtained from canonical quantization. Finally, we
show how these results are related.
\section{Stable forms in 7D}

In this section we study the relation between invariant stable forms and structures on a 7d Riemannian manifold, $\mathbb R^7$. To understand the geometric structures defined by stable forms, we need to study the isotropy subgroup of such forms under the action of the general linear group $GL(7)$. We start by recalling the structure on $\mathbb{R}^7$. Later we use such construction to understand the case of a manifold $X$.

Let $V$ be a real 7d vector space with basis $\{e_i\}$ and consider the space of $3$-forms $\wedge^3 V^*$. A form $\omega$ in $\wedge^3 V^*$ can be written as 
\begin{equation}
\omega=\sum_{i,j,k=1}^7 a_{ijk}e^{ijk},
\end{equation}
 where $e^{ijk}=e^i\wedge e^j\wedge e^k$ and $\{e^i\}$ is a basis for $V^*$. Consider the group $G=GL(7)$  of automorphisms of $V$. There is a natural action $G\curvearrowright \wedge^3V^*$ and it is known that there are two distinguished orbits given by this action, namely
\begin{align}
 G&\cdot \omega_1,\\
 G&\cdot \omega_2,
\end{align}
where $\omega_i$ is the form defined as
\begin{align}
\omega_1 & = e^{123} - e^{145} + e^{167} + e^{246} + e^{257} + e^{347} - e^{356}, \label{eq:w1} \\
\omega_2 & = e^{123} + e^{145} - e^{167} + e^{246} + e^{257} + e^{347} - e^{356}.\label{eq:w2}
\end{align}
To each form corresponds an isotropy group, the Lie group
\begin{equation}
 G_{\omega_1}=G_2,\quad
 G_{\omega_2}=\Tilde{G_2}.
\end{equation}
It is proved in \cite{bryant1987metrics} that $G_2$ is compact, connected, simple, simply connected, $14$-dimensional and it fixes the Euclidean metric $g_1=\sum (x^i)^2$ where $x= x^ie_i$ and $y= y^ie_i$ induced by
$$\langle x, y\rangle_{\omega_1} = x^1y^1+x^2y^2+x^3y^3+x^4y^4+x^5y^5+x^6y^6+x^7y^7.$$
$G_2$ also preserves the orientation of the forms $\omega_1$ and $*\omega_1$ with respect to $g_1$, and $G_2$ is isomorphic to the group of automorphisms of the octonians. 

There are analogous results for the group $\Tilde{G_2}$, this group preserves $\omega_2,*\omega_2$, the metric induced by
$$\langle x,y\rangle_{\omega_2}=x^1y^1+x^2y^2+x^3y^3-x^4y^4-x^5y^5-x^6y^6-x^7y^7,$$
and it is the non-compact dual of $G_2$. It is also connected, of dimension 14 and simple.

In this case the natural identification $$G\cdot \omega_i=G/G_{\omega_i},$$ is in fact a diffeomorphism. Since  $dim(G)=49$ and $dim(G_2)=dim(\Tilde{G_2})=14$ then the dimension of these orbits $dim(G\cdot \omega_i)=49-14=35$ coincides with the dimension of the ambient space $dim(\wedge^3V^*)=35$ and we conclude as in \cite{bryant1987metrics} that both orbits are open and the forms $\omega_1$ and $\omega_2$ are stable. In \cite{2007HVLe}, the authors show that the forms $\omega_1,\omega_2$ are essentially the unique stable forms, in the sense that any stable form $\omega\in \wedge^3V^*$ is either in the orbit of $\omega_1$ or $\omega_2$.

The scenario we study in this paper is the case when $X$ is a complete 7d Riemannian manifold, $x\in X$ is a point and $V=T_x X$. A stable form induces a $G_{\omega_i}$-structure on $X$, as follows (see \cite{Clarke2012HolonomyGI}) :

Consider the fiber bundle $\wedge^3T^*X$ and the open subbundle $\mathcal{P}^i(X)$ with fiber
$$\mathcal{P}^i_x=\{\omega\in  \wedge^3 V^*| \exists f:V\rightarrow \mathbb{R}^7\text{ with }f^*(\omega_i)=\omega\},$$
where in the last definition $f$ is an oriented isomorphism. From the previous discussion  $\mathcal{P}^3_x\cong G\cdot \omega_i$.
Fix a form $\omega$ over $X$ such that $\omega|_p\in\mathcal{P}^i_x=g\cdot\omega_i$ and consider the frame bundle $F$ of $X$ with fiber
$$F_x=\{f|f:V\rightarrow \mathbb{R}^7\text{ is an isometry}\}.$$
Let $Q$ be the principal subbundle of $F$ whose fiber consists in isomorphisms preserving $\omega$. Hence the fiber is $Q_x\cong G_{\omega_i}$ and $\omega$ determines $Q$ which defines a $G_{\omega_i}$-structure on $X$, preserving the metric $g_{\omega}$ induced by the inner product
$$\langle x,y\rangle_\omega=g\cdot\langle x,y\rangle_{\omega_i}.$$

There is a converse for this construction: given an oriented $G_{\omega_i}$-structure we can define a metric $g$, a $3$-form $\omega$ and $*\omega$ requiring that the corresponding metric is preserved by the action of $G_{\omega_i}$.

Let $X$ be a Riemannian 7d manifold with a $G_2$-structure $(\omega,g)$ and denote as $\nabla_g$ the Levi-Civita connection
associated to $g$. Let $\nabla_g\omega$ be the torsion of this $G_2$-structure. We say that $(\omega,g)$ is torsion-free if $\nabla_g\omega=0$. Finally define a $G_2$-manifold as a triplet $(X,\omega,g)$ such that $(\omega,g)$ is torsion-free.

Consider a $G_2$-manifold $X$. The existence of a $G_2$ holonomy metric is equivalent to the existence of a $3$-form $\Phi$ satisfying as in \cite{Dijkgraaf:2004te},
\begin{align}
\begin{split}
d\Phi&=0, \\
d_{*\Phi}\Phi &=0.
\end{split}
\end{align}
A stable $3$-form
can be written in terms of a 7d vielbein as 
\begin{equation}\Phi=\sum_{i,j,k=1}^7 \Psi_{ijk}e^i e^j e^k,\end{equation}
where $\Psi_{ijk}$ are the structure constants of the imaginary octonions. There are analogous constructions for stable forms on a $\Tilde{G_2}$-manifold, since the orbits of $\omega_1,\omega_2$ correspond with the holonomy groups $G_2$ and $\Tilde{G_2}$ respectively.

In order to define a volume on a $G_{\omega_i}$-manifold  $X$ consider a $3$-form $\Phi$ on $X$ as before, invariant by the corresponding holonomy group and define a volume as 
\begin{equation}
V_7(\Phi)=\int_X\Phi\wedge{}_{*\Phi}\Phi. \label{vform}
\end{equation}
As above since in the 7d case there are only two open orbits of maximal dimension hence is natural to consider only forms in these orbits to get a notion of {\it genericity} as in \cite{Dijkgraaf:2004te}.

\section{3D gravity from topological M-theory}
In \cite{Dijkgraaf:2004te}, Dijkgraaf et al. introduced a notion for TMT
in 7d with the property that it seems to unify several lower
dimensional topological models. In particular, they find a 
dimensional reduction that recovers  
the equations of motion of 2+1 gravity
from the volume of the 7d manifold $X$
discussed in the previous section.
A similar construction was given by Bryant et al.~\cite{bryant1987metrics}, where starting from a rank-4 spin bundle $\mathbf{S}$ over a 3d space of constant curvature (\textit{space form}), a 3-form $\Phi$
satisfying $d\Phi = d_{*\Phi}\Phi  = 0$ is constructed by making use of the structure equations for a manifold with constant sectional curvature $\kappa \equiv 4\Lambda$, i.e.,
\begin{subequations}\label{eqs:struc}
\begin{align}
    d e = - A\wedge e - e\wedge A\,, \\
    d A = - A\wedge A - \Lambda e\wedge e\,,
\end{align}
\end{subequations}
where $\{e^1, e^2, e^3\}$ is a basis of the tangent space at a point of the 3-manifold, and $A$
is a Levi-Civita connection 1-form. 
As~\cite{bryant1987metrics,Dijkgraaf:2004te} point out,
a 3-form that generalizes $\omega_1$~\eqref{eq:w1} can 
satisfy the conditions $d\Phi = d_{*\Phi}\Phi  = 0$ 
in some special cases. In order to write down this 3-form $\psi$ it is convenient to introduce first 
a set of local coordinates on the 4d fibre. Let $y_i$ be those coordinates, we define
$r=y_i y^i$. Notice that this is $SO(4)$-invariant. 
With the following 2-forms,
\begin{align}\label{eq:sigmalo}
\begin{split}
\Sigma^5&=e^{12}-e^{34}, \\
\Sigma^6&=e^{13}-e^{42}, \\
\Sigma^7&=e^{14}-e^{23},
\end{split}
\end{align}
we can write the 3-form $\Phi$ that satisfies $d\Phi = d_{*\Phi}\Phi  = 0$ 
as
\begin{equation}
    \Phi = f^3(r) e^{567} + f(r) g^2(r) e^m\wedge \Sigma^m\,.\label{eq:phiminus}
\end{equation}
Since $f$ and $g$ depend only on $r$, $\Phi$ preserves the $SO(4)$ invariance of $\omega_1$. Remembering that $SO(4)$
is a subgroup of $G_2$, and by the discussion of the previous section, the fact that $\Phi$ is 
$SO(4)$-invariant is a good indicator that it can define a $G_2$ structure -- thus satisfying the required equations.
The local coordinates $y_i$ are also used to define a 
basis of 1-forms in the fibre direction as
\begin{equation}
    \alpha = d y - yA\,.
\end{equation}
The four components of $\alpha$ are identified as a local basis on the fibre, $\alpha^i = e^i$, $i=4,5,6,7$. As a consequence of eqs.~\eqref{eqs:struc}, these 1-forms satisfy
\begin{equation}
    d\alpha = -\alpha\wedge A + (\kappa/4) y\omega\wedge\omega.\label{eq:dalpha}
\end{equation}
Using Eqs.~(\ref{eqs:struc}),(\ref{eq:dalpha}), and
\begin{equation}\label{phiast}
{}_{*\Phi}\Phi=-\frac{1}{6} g^4\Sigma_m\wedge\Sigma^m + \frac{1}{2}f^2 g^2\epsilon^{mnp}e^m\wedge e^n\wedge \Sigma^p,
\end{equation}
in ~\cite{0681.53021} it is showed that the equations $d\Phi = d_{*\Phi}\Phi  = 0$
hold if 
\begin{align}
\begin{split}
f(r)&=\sqrt{3\Lambda}(1+r)^{1/3}\,, \\
g(r)&=2(1+r)^{-1/6}\,.
\end{split}
\end{align}
Conversely, the authors of~\cite{Dijkgraaf:2004te} start with $d\Phi = d_{*\Phi}\Phi  = 0$ and verify that the above assumptions for $f(r)$ and $g(r)$
lead to the structure equations,~(\ref{eqs:struc}), i.e., in their interpretation, the equations of motion for 3d gravity arise from the equations for a 3-form with $G_2$-holonomy. 
If these equations of motion are recovered from such a 3-form $\Phi$, it is natural to look for a Lagrangian for $\Phi$
that encompasses the main points of the derivations above and reduces to the known Lagrangians for 3d gravity. This Lagrangian is given precisely in terms of the volume form discussed around Eq.~(\ref{vform}). In order to convert~Eq.~(\ref{vform}) into an expression
that we can recognise as the action for 2+1 gravity we perform the following steps. First, we rewrite the integrand $\Phi\wedge{}_{*\Phi}\Phi$ using the antisymmetry  
of the wedge product and of the Levi-Civita tensor, obtaining
\begin{equation}
V_7(\Phi)=\int_X \frac{40}{3}(3\Lambda)^{3/2}(1+r)^{1/3}e^{567}\wedge\Sigma_i\wedge\Sigma^i\,.
\end{equation}
Now, let $\Sigma$ be the curvature of a connection
$\alpha$, i.e., 
\begin{equation}
\Sigma_5=d\alpha_5 + 2\alpha_6 \alpha_7\,,
\label{eq:sigmacurv}
\end{equation}
and cyclically for the others. Later
on we will relate this $\alpha$ to the connection 1-form $A$. Notice that this is compatible with the
equations~\eqref{eq:sigmalo} that
express $\Sigma^i$ in a local orthonormal basis~\cite{hitchin2001stable}.
Using again the properties of the wedge product, and noticing that as a consequence of the structure equations~\eqref{eqs:struc} we have
$d(e^{567}) = 0$~\cite{0681.53021},
the volume $V_7$ can be written as
\begin{align}
V_7(\Phi)=\int_X & \frac{40}{3}(3\Lambda)^{3/2}(1+r)^{1/3}d\left[e^{567}\wedge(\alpha_i\wedge d\alpha_i \right. \nonumber \\
& \left.+\frac{2}{3} \epsilon^{ijk}\alpha_i\alpha_j\alpha_k) \right]\,.
\end{align}
The argument of the differential does not depend on $r$, therefore, by an appropriate choice of coordinates, its prefactor can be integrated out so that it becomes a global factor of a 6d integral. We can further reduce these dimensions by using Stokes theorem, obtaining\footnote{We have to be careful with the notation: all $p$-forms
are integrated over $p$-dimensional manifolds. If the dimensions of the integral and the order of the $p$-form obtained by counting wedge products does not match, this means that one of the differentials $dx^i$ has been integrated out, and we have to remember this when writing the form in component notation.}
\begin{align}
V_7(\Phi)\propto\int_{X^5} & e^{567}\wedge(\alpha_i\wedge d\alpha_i +\frac{2}{3} \epsilon^{ijk}\alpha_i\alpha_j\alpha_k) \,.
\end{align}
Finally, since the argument of the integral only depends on quantities defined over the 3-manifold $\mathcal M$ with basis $\{e^5,e^6,e^7\}$, the
volume can be expressed as
\begin{equation}
 V_7(\Phi)\sim \int_{\mathcal M} e^{567}\wedge(\alpha_i\wedge d\alpha_i+\frac{2}{3} \epsilon^{ijk}\alpha_i\alpha_j\alpha_k) .
\end{equation}
Expanding the wedge product in components, relabeling the internal indice as $(a,b,c)$ and using $(i,j,k)$ for the spacetime indices, we get
\begin{equation}
 V_7(\Phi)\sim \int_{\mathcal M} \epsilon^{ijk}(2\alpha^a_i\wedge \partial_j\alpha^a_k+\frac{2}{3} \epsilon_{abc}\alpha^a_i\alpha^b_j\alpha^c_k) .
\end{equation}
This is the Chern-Simons action. At this point it is convenient to
notice that the 2-forms $\Sigma$ are anti-self-dual, i.e., $^*\Sigma^i = -\Sigma^i$. For this reason, we rename it as $^-\Sigma^i$, with associated connection $^-\alpha_i$, and we also rename the form $\Phi$ given in eq.~\eqref{eq:phiminus} as $^-\Phi$. Now we are ready to see the relevance of the discussion of the previous section. The form $^-\Phi$
is constructed out of the stable form
$\omega_2$ presented in eq.~\eqref{eq:w2}. However, we have seen that
the volume form can also be constructed in terms of $\omega_1$, eq.~\eqref{eq:w1}. Furthermore, these two possibilites, $\omega_1$ and $\omega_2$ are unique in the sense discussed in the previous section. With these considerations in mind, we construct a volume form for each of the 3-forms
\begin{align}
    ^-\Phi &=  f^3(r) e^{567} + f(r) g^2(r) e^m\wedge {}^-\Sigma^m\,,\label{eq:phiminus2}\\
    {}^+\Phi &=  f^3(r) e^{567} + f(r) g^2(r) e^m\wedge {}^+\Sigma^m\,.\label{eq:phiplus2}\, , 
\end{align}
where ${}^+\Sigma^m$ are the self-dual 2-forms
\begin{align}
\begin{split}
^+\Sigma^5&=e^{12}+e^{34}, \\
^+\Sigma^6&=e^{13}+e^{42}, \\
^+\Sigma^7&=e^{14}+e^{23},
\end{split}
\end{align}
and $r$ is defined in the same way as described before. 
When $f(r)=g(r)=1$, ${}^- \Phi$, ${}^+ \Phi$ are equivalent to $\omega_2$ and $\omega_1$, respectively. The 4-forms associated to $^-\Phi$ and
$^+\Phi$ are
\begin{align}\label{phiastminusplus}
{}_{*\Phi}{^\mp\Phi}=& \mp\frac{1}{6} g^4{}^\mp\Sigma_m\wedge{}^\mp\Sigma^m \nonumber \\ 
& \pm \frac{1}{2}f^2 g^2\epsilon^{mnp}e^m\wedge e^n\wedge {}^\mp\Sigma^p\,.
\end{align}
We can use either of ${}^\pm\Phi$ to construct the volume of the 7-manifold $X$,
\begin{equation}
 V^{\pm}\equiv   V_7({}^\pm\Phi) = \int_X {}^\pm\Phi\wedge {}_{*\Phi}{^\pm\Phi}\,.
\end{equation}
By the same steps of the previous section, $V_7$ can be written as
    \begin{equation}
 V^{\pm}\sim \int_{\mathcal M} \epsilon^{ijk}(2{}^\pm\alpha^a_i\wedge \partial_j{}^\pm\alpha^a_k+\frac{2}{3} \epsilon_{abc}{}^\pm\alpha^a_i{}^\pm\alpha^b_j{}^\pm\alpha^c_k) ,
\label{eq:cspm}
\end{equation}
where ${}^+\alpha^i$ is the connection associated to ${}^+\Sigma^i$.
Thus, we have found two Chern-Simons actions
derivable from the volume of a 7-manifold that admits
two special stable forms. Now we want to understand how these
two actions are related to 2+1 gravity. From the results of
~\cite{0681.53021,Dijkgraaf:2004te}, we know that
the equations of motion arising from the volume of ${}^-\Phi$  are those of 2+1 gravity with a cosmological constant. Since $V({}^+\Phi)$
describes the same volume as $V({}^-\Phi)$, the 3d equations of motion
derived from both actions have to coincide. This is remarkably similar, 
and consistent, with the results of~\cite{Witten:1988hc}, where
it is shown that there are two 3d actions, named
\textit{standard} and \textit{exotic}, that lead 
to the same equations of motion that we are interested in. Furthermore,
they show that these actions can be written precisely 
in terms of the Chern-Simons actions~\eqref{eq:cspm} by setting
\begin{equation}
    {}^\pm\alpha^a_i = A_i^a \pm \sqrt{\lambda}e^a_i\,,
\end{equation}
where $A_i$ and $e_i$ are the fields introduced around Eq.~\eqref{eqs:struc}. 
The combinations
\begin{align}
I_{st}&=\frac{^+I - ^- I}{4\sqrt{\lambda}}\,,  \label{eq:ist} \\
I_{ex}&=\frac{^+I + ^- I}{2},  \label{recoverstex}
\end{align}
where ${}^\pm I$ are the integrals in Eq.~(\ref{eq:cspm}), give respectively the standard and exotic actions. 

Now we can reinterpret the standard and exotic actions in terms of  the volume functional as
\begin{align}
I_{st}&=\frac{h^+ V^+ - h^- V^-}{4\sqrt{\lambda}}  \nonumber \\
I_{ex}&=\frac{h^+ V^+ + h^-V ^- }{2}.
\end{align}
where $h^{\pm}$ are the inverses of the proportionality factors in Eq.~\eqref{eq:cspm}. In this way, we can see the standard and exotic actions as two different combinations of pieces of the volume of the 7-manifold X. Applications of the ideas developed so far to the Immirzi ambiguity in 3d gravity have been presented in~\cite{Chagoya_2018}.
In the next section we explore the entropy of the BTZ black hole
from the point of view of TMT and we discuss the
relation of our results to the conjecture $Z_{BH} = |Z_{top}|^2$. 
\section{BTZ black hole: partition function}

Using the results described above we can provide evidence that 
the conjecture discussed around Eq.~\eqref{eq:ooguri}
also applies for $G_2$-manifolds and 3d black holes, i.e., that in general, the partition function of a theory with action 
defined by a Hitchin functional is related to the partition function for a BPS black hole in
the gravitational theory allowed by the $p$~-forms used to construct the Hitchin functional.
The possibility that the relation between BPS objects and form
theories of gravity extends to G2-manifolds was hinted in~\cite{Dijkgraaf:2004te}; however, it was only studied for 4d and 5d black holes embedded in a 6d $SU(3)$-manifold.
In this work we show explicitly that the partition function of the BTZ black hole is recovered from the partition function associated to the
volume $V_7$. Given the different ways of
writing down $V_7$ either in terms of $V^+, V^-$ or both,
one could think that the result only applies to the extremal case, which turns out to
be associated to the situation were we demand that the linear combinations of $V^+$ and $V^-$ -- for instance $I_{st}$ and $I_{ex}$ --
preserve  a given multiple of $V_7$; but as we argue below, the partition function obtained from TMT correctly gives the
BH partition function even away from the extremal case.

In the case of TMT, the total space $X$ is 7d and as we shown in the previous sections, its volume can be 
constructed with either of the 3-forms 
${}^+\Phi$ and ${}^-\Phi$. 
A certain combination of these volumes, Eq.~(\ref{eq:ist}), results in the standard
action for 3d gravity. In this theory,
a black hole solution
is given by the BTZ space-time \cite{Banados1992}, whose metric can be written as 
\begin{equation}
ds^2 = -N^2 dt^2 + N^{-2} dr^2 + r^2(N^\phi dt + d\phi)^2,
\end{equation}  
where the lapse $N$ and shift $N^\phi$ are
\begin{align}
N &= \left(-M + \frac{r^2}{{\ell}^2} + \frac{J^2}{ 4 r^2} \right)^{1/2}, \\
N^\phi & = -\frac{J}{2 r^2}.
\end{align}
The integration constants $M$ and $J$ are interpreted respectively as the mass and angular 
momentum of the black hole, and $\ell$ is related to the cosmological constant of the theory by 
$\ell^{-2} =\Lambda/3$. The lapse function vanishes at two distinct values of $r$, defining
two coordinate singularities, $r_{\pm}$, 
\begin{equation}
    r_{\pm} = \frac12 \left(  \sqrt{\ell (\ell M+J) }  \pm \sqrt{\ell(\ell M-J)}\right)\,.
\end{equation}
When $J=0$ only $r_+$ is different from zero, and in the
extremal case $J=M\ell$ the two horizons coincide. The entropy of the BTZ black hole can
be computed by different methods, for example, by Euclidean path integral or by Noether
charges \cite[see e.g.][]{CarlipClass.Quant.Grav.12:2853-28801995}, and it is given by
\begin{equation}
S_{BTZ}^{st} = 4\pi r_+.\label{smas}
\end{equation}
These computations do not depend only on the metric but also on the action, that is usually taken to be the standard action, hence
the superscript $st$. Originally, this result comes from geometrical considerations on the standard action of 2+1 gravity, and then deriving the entropy from the grand canonical partition function in the classical approximation~\cite{Banados:1993qp} 
$$
Z = \exp(I_{st})\,.
$$
 Since the standard action is recovered from TMT, the entropy of a BTZ black hole described by such an action is recovered as well.

The same techniques that lead to Eq.~\eqref{smas} have been applied to the exotic action, finding an entropy proportional
to the inner BTZ horizon, $r_-$. The fact that the entropy is
proportional to the inner horizon raised doubts about the
validity of black hole thermodynamics. However, it has been shown
that these laws hold~\cite{Townsend2013}. Indeed, the result is even more general: an entropy of the form 
\begin{equation}S\sim \alpha r_+ + \gamma r_-
\label{eq:mixen}
\end{equation}
is in agreement with black hole thermodynamics. 
%
%
%
Eq.~\eqref{eq:mixen} arises naturally in the
context we are studying in this work.
%
Hitchin's partition function is defined in terms of the volume
functional,
\begin{equation}
    Z_H(\Phi) = \int_{[\Phi]} d\Phi \,\textrm{exp}{(V_H(\Phi))}\,.
\end{equation}
Thus, when we write TMT as a theory of a 4d 
vector bundle over a 3d base space such that the 7d manifold $X$ has a $G_2$-structure, we can separate $V_7$ in terms of the volume functionals $V^{\pm}$,
\begin{equation}
   \lambda V_7 = \beta_+ V^+ + \beta_- V^- \,,\label{eq:volsplit}
\end{equation}
for some coefficients $\lambda, \beta_{\pm}$. Notice that, so far, all the properties that hold for a theory based on $V_7$ hold for a theory based on a multiple $\lambda$ of $V_7$. 
In addition, $V^{\pm}$ are proportional to the Chern-Simons actions,
 Eq.~\eqref{eq:cspm}, with proportionality constants $1/h^\pm$.
Putting all together, we write Hitchin's partition function as
\begin{equation}
Z_H(\Phi) = \int_{[\Phi]} d\Phi \, \textrm{exp}\left[\sum_{\sigma=+,-} \beta_\sigma(h^{\sigma})^{-1}\, {}^\sigma I\right],
\end{equation} 
As before, the basis of the 7d manifold can be decomposed into a 3d
base space and a 4d bundle. 
  The coefficients
$\beta_\pm$ can be chosen in such a way that the linear combination of $^\pm I$ in the argument of the exponential reproduces either the 
standard or the exotic action, or a combination of both. For the choice that leads to the standard action, by the discussion above we confirm  that Hitchin's entropy is related to the BTZ entropy,
\begin{equation}
Z_H(\Phi) \propto \int de d\alpha\, \textrm{exp}(I_{st})  = Z_{BH}\,.
\label{entropy}
\end{equation}
On the other hand, for a different choice of parameters we can have
\begin{equation}
Z_H(\Phi) \propto \int de d\alpha\, \textrm{exp}(I_{ex} ) = Z_{BH} \,,
\label{entropy2}
\end{equation}
i.e., the Hitchin partition function for the exotic action is also related to a black hole partition function, only that in this case $Z_{BH}$ corresponds to the exotic BTZ black hole.

The extremal case, $r_+ = r_-$, admits an interpretation
from the point of view of TMT. Suppose we fix $\lambda$, e.g. $\lambda=1$. This imposes a constraint
on the linear combinations in Eq.~\eqref{eq:volsplit}, such
that any choice of $\beta_\pm$ leads to a fixed $V_7$ and the
same $Z_H(\Phi)$. Therefore, all combinations lead to 
the same black hole entropy, and this is only possible 
if $r_+ = r_-$, i.e., the extremal case corresponds to
a constraint on the parameters $\beta_{\pm}$.
\section{Discussion}
3d gravity can be embedded in a 7-manifold
with $G_2$-holonomy. The volume form of this manifold
is constructed in terms of a stable (\textit{generic}, in the sense of~\cite{Dijkgraaf:2004te}) form. Indeed, there
are essentially two unique such forms and by using these two 
stable forms, we split the volume of the 7-manifold into contributions from the distinct orbits. Using the
structure equations appropriated for our geometrical set-up,
we find that these two contributions can be rephrased as
Chern-Simons actions, one for a self-dual curvature and one for an anti-self-dual curvature. This observation allows us
to recover the two classically equivalent known actions 
of 3d gravity, i.e., Witten's standard and exotic
actions, thus completing the picture shown in \cite{0681.53021,Dijkgraaf:2004te}. 

In a context that is more general than the theory that we study here, it has been conjectured that topological and black hole partition functions are related. Our results give a 
concrete realisation of this conjecture: by writing the action of TMT in terms of the
contributions from the two unique stable forms, we can tune
the theory so that it reproduces the partition function
of the standard action of 3d gravity, thus
agreeing with the result for the BTZ black hole; or we
can choose to reproduce the exotic action, obtaining the correct entropy for the exotic BTZ black hole. It is worth noticing that a combined standard/exotic entropy is in agreement with black hole thermodynamics~\cite{Townsend2013}, and our results provide a scenario where such combined
models can be embedded. 

The topological partition function is also conjectured to be 
related to a wave function. The wave function for a static BTZ black hole in the region
outside the horizon has been computed within a canonical quantization scheme~\cite{Vaz2008}. When evaluated at the horizon, their result takes the form (more details in the Appendix):
$$
|\psi|^2 \sim e^{\tilde \mu r_+} \,,
$$
where $\tilde \mu$ is a quantized number related to the 
energy levels of the system. This result indeed reassembles
the Euclidean partition function for the BTZ black hole. It would be interesting to
explore the quantization of a non-static BTZ black hole, so that the relation between the wave function and the 
black hole partition function can be explored for the extremal case, i.e., the case that would correspond to the 
conjectures in~\cite{Ooguri:2004zv}. This is left for future work.

\section*{Acknowledgments}
This work is  supported by   CONACYT grants  257919, 258982. M. S. is supported by CIIC 28/2020.

\bibliographystyle{unsrt}
\bibliography{ref}

\begin{thebibliography}{10}

\bibitem{Witten:1988hc}
Edward Witten.
\newblock {(2+1)-Dimensional Gravity as an Exactly Soluble System}.
\newblock {\em Nucl.Phys.}, B311:46, 1988.

\bibitem{Banados1992}
M\'aximo Ba{n}ados, Claudio Teitelboim, and Jorge Zanelli.
\newblock The black hole in three dimensional space time.
\newblock {\em Phys.Rev.Lett.}, 69:1849--1851, 1992.

\bibitem{Townsend2013}
Paul~K. Townsend and Baocheng Zhang.
\newblock {Thermodynamics of Exotic Ba\~nados-Teitelboim-Zanelli Black Holes}.
\newblock {\em Phys. Rev. Lett.}, 110(24):241302, 2013.

\bibitem{PhysRevD.61.085022}
H.~Garc\'ia-Compe\'an, O.~Obreg\'on, C.~Ram\'irez, and M.~Sabido.
\newblock Remarks on 2+1 self-dual chern-simons gravity.
\newblock {\em Phys. Rev. D}, 61:085022, Mar 2000.

\bibitem{Dijkgraaf:2004te}
Robbert Dijkgraaf, Sergei Gukov, Andrew Neitzke, and Cumrun Vafa.
\newblock {Topological M-theory as unification of form theories of gravity}.
\newblock {\em Adv.Theor.Math.Phys.}, 9:603--665, 2005.

\bibitem{2007HVLe}
H.-V. {Le}, M.~{Panak}, and J.~{Vanzura}.
\newblock {Manifolds admitting stable forms}.
\newblock {\em ArXiv e-prints}, April 2007.

\bibitem{Chagoya_2018}
Javier Chagoya and Miguel Sabido.
\newblock Topological m-theory, self-dual gravity and the immirzi parameter.
\newblock {\em Classical and Quantum Gravity}, 35(16):165002, jul 2018.

\bibitem{Ooguri:2004zv}
Hirosi Ooguri, Andrew Strominger, and Cumrun Vafa.
\newblock {Black hole attractors and the topological string}.
\newblock {\em Phys. Rev. D}, 70:106007, 2004.

\bibitem{witten1993quantum}
Edward Witten.
\newblock Quantum background independence in string theory, 1993.

\bibitem{Vaz2008}
Cenalo Vaz, Sashideep Gutti, Claus Kiefer, T.~P. Singh, and L.~C.~R.
  Wijewardhana.
\newblock {Mass spectrum and statistical entropy of the BTZ black hole from
  canonical quantum gravity}.
\newblock {\em Phys. Rev.}, D77:064021, 2008.

\bibitem{bryant1987metrics}
Robert~L Bryant.
\newblock Metrics with exceptional holonomy.
\newblock {\em Annals of mathematics}, 126(3):525--576, 1987.

\bibitem{Clarke2012HolonomyGI}
Andrew Clarke and Bianca Santoro.
\newblock Holonomy groups in riemannian geometry.
\newblock {\em ArXiv e-prints}, 2012.

\bibitem{0681.53021}
Robert~L. Bryant and Simon~M. Salamon.
\newblock {On the construction of some complete metrics with exceptional
  holonomy.}
\newblock {\em Duke Math. J.}, 58(3):829--850, 1989.

\bibitem{hitchin2001stable}
Nigel Hitchin.
\newblock Stable forms and special metrics, 2001.

\bibitem{CarlipClass.Quant.Grav.12:2853-28801995}
Steven Carlip.
\newblock The (2+1)-dimensional black hole.
\newblock {\em Class.Quant.Grav.}, 12:2853-2880, 1995.

\bibitem{Banados:1993qp}
Maximo Banados, Claudio Teitelboim, and Jorge Zanelli.
\newblock {Black hole entropy and the dimensional continuation of the
  Gauss-Bonnet theorem}.
\newblock {\em Phys. Rev. Lett.}, 72:957--960, 1994.

\bibitem{Strominger:1997eq}
Andrew Strominger.
\newblock {Black hole entropy from near horizon microstates}.
\newblock {\em JHEP}, 02:009, 1998.

\end{thebibliography}

\appendix*
\section{Stationary states of the BTZ Black Hole}\label{app}
{In this appendix we shortly present the derivation done in \cite{Vaz2008}, of the non rotating BTZ black hole  wave function 
\begin{equation}
    \Psi=e^{(i/4G)\int_0^\infty dr\Gamma(r)W(\tau(r),R(r),F(r))},
\end{equation}
where $\tau=\tau(0), R=R(0)$ and $F=F(0)$. The WDW equation becomes the KG equation
\begin{equation}
    \left[ \frac{\partial^2}{\partial\tau^2}+ F \frac{\partial^2}{\partial R^2} + A \frac{\partial^2}{\partial R}+B\right]e^{i\mu W(\tau,R,F)}=0.
\end{equation}
As these description is based on a collapsing shell we impose that we have a free wave function, this is a natural assumption. For this to hapen we should be able to write the WDW equation as
\begin{equation}
    \gamma^{ab}\nabla_a\nabla_b\Psi=0,
\end{equation}
where $\gamma^{ab}$ is the DeWitt supermetric on the configuration space and $\nabla_a$ is the covariant derivative.The WDW equation is the free KG equation if $B=0$ and $A(R,F)=\vert F\vert\partial_R \ln{\sqrt{\vert F\vert}}$ and the inner product is given by
\begin{equation}
    <\Psi_1,\Psi_2>=\int\frac{dR}{\sqrt{\vert F\vert}}\Psi_1^*\Psi_2
\end{equation}
when $F\ne0$, the supermetric can be written in a flat form by the transformation $R_*=\pm\int \vert R\vert^{-1/2}dR$. In terms of $R_*$ the KG equation is
\begin{equation}
    \left[ \frac{\partial^2}{\partial\tau^2}\pm \frac{\partial^2}{\partial R_*^2}\right]e^{i\mu W(\tau,R,F)}=0,
\end{equation}
the positive sign is for the exterior and the minus sign for the interior.
The solutions are
\begin{equation}
    \begin{array}{cc}
        \psi^{in}(\tau,R_*)=A_\pm e^{-i\mu (\tau\pm R_*)} &  F<0,\\
        \psi^{out}(\tau,R_*)=B_\pm e^{-i\mu (\tau\pm iR_*)}  & F>0.
    \end{array}
\end{equation}
In the exterior 
\begin{equation}
    R_*=\frac{1}{\sqrt{\Lambda}}\left [\ln{\left( \frac{R\sqrt{\Lambda}+\sqrt{\Lambda R^2-8GM}}{\sqrt{8GM}}\right )}+\frac{\pi}{2}\right],
\end{equation}
at the horizon $R_*=\ln{r_+}$.
For a continuous wave function the matching conditions give the following spectrum
\begin{equation}
\mu_j=\sqrt{\Lambda}\hbar\left ( j+\frac{1}{2}\right ),\quad j=0,1,2,\dots
\end{equation}

a similar spectrum was derived in \cite{Strominger:1997eq}.}
 \end{document}